\documentclass[conference]{IEEEtran}
\IEEEoverridecommandlockouts
\usepackage{cite}
\usepackage{amsmath,amssymb,amsfonts}
\usepackage{algorithmic}
\usepackage{graphicx}
\usepackage{textcomp}
\usepackage{xcolor}

\usepackage[T1]{fontenc}
\usepackage{helvet}

\usepackage{caption}
\usepackage{subcaption}

\def\BibTeX{{\rm B\kern-.05em{\sc i\kern-.025em b}\kern-.08em
    T\kern-.1667em\lower.7ex\hbox{E}\kern-.125emX}}
\begin{document}

\title{Superposition of PRS and PDSCH for ISAC System: Spectral Efficiency Enhancement and Range Ambiguity Elimination\\
\thanks{This work was funded by the European Commission through the H2020 MSCA 5GSmartFact project under grant agreement number 956670.}
}

\author{Keivan Khosroshahi\IEEEauthorrefmark{1}\IEEEauthorrefmark{2}, Philippe Sehier\IEEEauthorrefmark{2}, and Sami Mekki\IEEEauthorrefmark{2},\\
\IEEEauthorrefmark{1}
Université Paris-Saclay, CNRS, CentraleSupélec, Laboratoire des Signaux et Systèmes, Gif-sur-Yvette, France\\
\IEEEauthorrefmark{2} Nokia Standards, Massy, France
\\ keivan.khosroshahi@centralesupelec.fr, \{philippe.sehier, sami.mekki\}@nokia.com
}

\maketitle
 
\begin{abstract}
From the telecommunication companies' perspective, the preference for integrated sensing and communication (ISAC) for sixth-generation (6G) is to enhance existing infrastructure with sensing capabilities while minimizing network alterations and optimizing available resources. This prompts the investigation of ISAC leveraging the existing infrastructure of fifth-generation (5G) new radio (NR) signals as defined by the 3rd generation partnership project (3GPP).  Additionally, improving spectral efficiency is crucial in scenarios with high demand for both communication and sensing applications to maintain the required quality of service (QoS). To address these challenges, we propose the superposition of the physical downlink shared channel (PDSCH) for communication and the positioning reference signal (PRS) for sensing with proper power allocation. Furthermore, we propose a novel algorithm to reduce the interference for data decoding caused by PRS. Moreover, we introduce the joint exploitation of PRS and demodulation reference signal (DMRS) to prevent range ambiguity in the form of ghost targets. Through simulation analysis, we demonstrate the effectiveness of integrating PDSCH and PRS symbols within a unified resource grid.  Our results show that the introduced approaches not only eliminate range ambiguity when sensing targets from gNBs but also enhance spectral efficiency by reducing interference between PRS and PDSCH. Simulation results show throughput enhancement and up to $57\%$ improvement in bit error rate (BER). This paves the way for supporting sensing applications in the forthcoming network generation. \\
\end{abstract}


\section{Introduction}
The incorporation of integrated sensing and communication (ISAC) technology into forthcoming mobile networks is gaining widespread, driven by the anticipation of future applications such as intrusion detection, health monitoring, the internet of things (IoT), autonomous driving, etc., which impose novel demands on wireless communication networks \cite{zhang2021enabling}. Combining communication and sensing holds promise for enhancing spectral and energy efficiency while reducing hardware costs. There is increasing interest in developing ISAC signals based on reference signals due to their superior passive detection performance, robust anti-noise capabilities, and favorable auto-correlation characteristics \cite{wei20225g}. Among the reference signals available in fifth-generation (5G) networks, the positioning reference signal (PRS) stands out for sensing purposes due to its rich time-frequency resources and flexible configuration. Introduced in 3rd generation partnership project (3GPP) Release 16 of the 5G specification, PRS aims to enhance the position accuracy of connected user equipment (UEs) because of its high resource element (RE) density and superior correlation properties compared to existing reference signals \cite{3gpp2018nr3}. On the other hand, in 5G new radio (NR), the physical downlink shared channel (PDSCH) serves as the physical downlink channel for user data transmission, with the demodulation reference signal (DMRS) serving as the associated reference signal. DMRS, replacing the cell-specific reference signal (CRS) in LTE, aids in channel estimation and PDSCH decoding \cite{3gpp2018nr}. In the ISAC framework, where communication and sensing co-occur, PDSCH can be jointly utilized for data communication alongside PRS for sensing. 
As for the research focused on using reference signals for sensing, PRS is utilized in \cite{wei20225g} for range and Doppler estimation, with its performance compared to other reference signals specified in 5G. 
In \cite{huang2022joint}, a two-stage scheme for joint pilot optimization, target detection, and channel estimation is introduced for the ISAC scenario.
Additionally, in \cite{ma2022downlink}, DMRS and channel state information reference signal (CSI-RS) are separately used for velocity and range estimation. The authors in \cite{wang2020multi} investigate a multi-range joint automotive radar and communication system based on orthogonal frequency division multiplexing (OFDM) waveform. In \cite{khosroshahi2024dopplerambiguityeliminationusing}, the authors concentrate on Doppler ambiguity.

In sensing applications, the structure of active subcarriers in PRS can introduce ambiguity in range estimation, leading to ghost targets. The solution proposed in \cite{wei2023multiple} addresses this issue by using multiple reference signals to mitigate range ambiguity. However, it does not consider the interference between PDSCH used for data transmission and other reference signals such as PRS and CSI-RS. Additionally, the proposed configuration lacks flexibility in selecting different comb sizes. The method discussed in \cite{wei20225g} significantly reduces the maximum detection range, especially when the PRS comb size is large. The approach outlined in \cite{khosroshahi2024leveragingprspdschintegrated} is limited to the number of PRS slots. Furthermore, the irregular arrangement of PRS symbols proposed in \cite{brunner2024bistaticofdmbasedisacovertheair} and \cite{golzadeh2024joint} increases the noise floor and degrades the signal-to-noise ratio (SNR). The solution introduced in \cite{mura2024optimizedwaveformdesignofdmbased} imposes high computational complexity on the system and does not take 3GPP constraints into account. Additionally, resource management becomes critical in scenarios with high demand for both communication and sensing. Therefore, to the best of our knowledge, a 3GPP-compliant solution that enhances spectral efficiency and sensing coverage without resulting in ghost targets within the ISAC framework has yet to be proposed.

This paper proposes the superposition of PDSCH for data transmission and the PRS for sensing within an OFDM resource grid while the sensing coverage range is preserved without encountering ghost targets. This approach enhances spectral efficiency by fully utilizing resources in both time and frequency domains for communication and sensing, leading to high sensing resolution.
To address interference between PRS and PDSCH symbols sharing the same REs, we develop an algorithm to improve communication performance. Additionally, to eliminate ghost targets in range estimation, we introduce two algorithms for different PRS comb sizes and reuse the DMRS for sensing alongside data decoding. Our simulation results demonstrate the elimination of ghost targets with appropriate power allocation using the proposed algorithm. Furthermore, we show that the introduced interference cancellation method can increase throughput and reduce the bit error rate (BER) by up to $57\%$.

The rest of the paper is organized as follows: Section \ref{Sec2} provides a brief background on the 5G NR reference signals used in this work. In Section \ref{Sec3}, we detail the system model and the method for estimating the range and velocity of targets, along with an explanation of the range ambiguity problem. The steps for interference cancellation and range ambiguity elimination are presented in Section \ref{Sec5}. Section \ref{results} evaluates the performance of the proposed algorithms. Finally, Section \ref{Sec6} provides the concluding remarks and future works.

\section{PRS and DMRS as Reference Signals for Sensing} \label{Sec2}
The allocation of physical resource blocks (PRBs) in 5G NR, as defined in the technical specification (TS) 38.214 \cite{3gpp2018nr1}, is based on the concept of a PRB consisting of 12 consecutive subcarriers in the frequency domain and $14$ symbols in the time domain. Based on TS 38.211 \cite{3gpp2018nr}, the generation of the PRS and DMRS sequences can be done according to the following equation

\vspace{-3mm}
\begin{equation}
   \psi(m) = \frac{1}{\sqrt{2}}(1 - 2~c(2m)) + j\frac{1}{\sqrt{2}}(1 - 2~c(2m + 1))
   \label{r}
\end{equation}
where $c(i)$ denoted as Gold sequence of length-$31$. The starting value of $c(i)$ for the PRS and DMRS is provided in \cite{3gpp2018nr}.
PRS allocation encompasses a minimum of 24 PRBs and a maximum of 272 PRBs that shows the flexible transmission parameters supported in 5G NR. This flexibility enables PRS to adapt its time-frequency resource configuration to meet sensing accuracy requirements across diverse application scenarios. As per the PRS resource mapping guidelines indicated in TS 38.211 \cite{3gpp2018nr}, four comb structures—Comb 2/4/6/12—in the frequency domain and five symbol configurations—Symbol 1/2/4/6/12—in the time domain is supported.

DMRS generation occurs within the PDSCH allocation, as explained in TS 38.211 \cite{3gpp2018nr}. The resources allocated for PDSCH operation are located within the Bandwidth Part (BWP) of the carrier, as specified in TS 38.214 \cite{3gpp2018nr1}. DMRS plays a pivotal role in channel estimation and is placed within the resource blocks (RBs) assigned for PDSCH. The structure of DMRS is designed to accommodate various deployment scenarios and use cases. The positioning of DMRS symbols depends on the mapping type, which can be either slot-wise (Type A) or non-slot-wise (Type B). The positions of any additional DMRS symbols are determined by a set of tables, as outlined in TS 38.211 \cite{3gpp2018nr}. Moreover, 1 to 4 OFDM symbols in the time domain can be occupied by front-load DMRS. In this work, we reuse DMRS for sensing besides communication.

\begin{figure}[!t]
\centering
\mbox{\includegraphics[width=\linewidth]{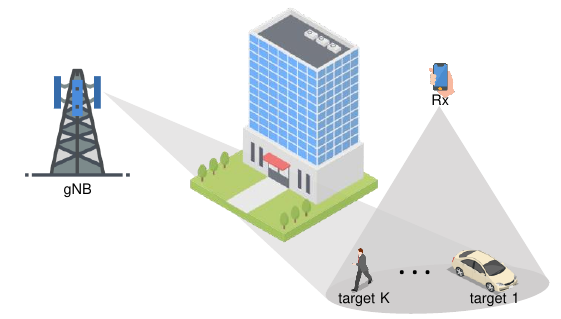}}
\caption{ISAC scenario where the UE acts as the simultaneous sensing and communication receiver.}
\label{bistatic RIS2}
\end{figure}

\section{ISAC System Model} \label{Sec3}
The current architectures of 5G NR signals were not designed for radar sensing. Thus, in this section, we propose a methodology for effectively leveraging both communication and sensing signals within the ISAC framework.
We assume a downlink ISAC scenario in which a gNB transmits the signal toward a scanning area containing $K$ point-like targets. Subsequently, the receiver not only captures echoes from these targets to estimate their range and velocity but also receives the communication signals. We assume the targets are in the line-of-sight (LoS) of both the transmitter and the receiver. An example of the scenario can be seen in Figure \ref{bistatic RIS2}. 

We consider that the OFDM resource grid comprises $N$ symbols in the time domain and $M$ subcarriers in the frequency domain. We define $S = \frac{N}{14}$ as the number of slots in a resource grid. Then, the transmitted time-domain signal can be represented as follows \cite{buzzi2019using}

\begin{align}
    x(t) =& \sum_{n = 0}^{N-1}\text{rect}(\frac{t - nT_0}{T_0})\sum_{m=0}^{M-1} (\sqrt{\gamma_c}  s_c(m,n) \notag \\
    &+ \sqrt{\gamma_s} s_s(m,n)) e^{j2\pi m\Delta f(t - nT_0)}
\end{align}
where $\text{rect}(t/T_0)$ denotes the rectangular pulse, $\Delta f$ represents the subcarrier spacing, and $\gamma_c$ and $\gamma_s$ are power allocation weights for communication and sensing, respectively, with $\gamma_s + \gamma_c = 1$. $T_0 = T_{CP} + T_s$ is the total duration of the OFDM symbol, where $T_{s} = \frac{1}{\Delta f}$ represents the symbol duration, and $T_{CP}$ denotes the duration of the cyclic prefix (CP). $s_c(m,n)$ and $s_s(m,n)$ correspond to the modulated communication and sensing symbols, respectively, where $n=0,...,N-1$ and $m=0,...,M-1$ within an $M \times N$ OFDM resource grid. The signals reflected and received by the receiver can be represented as \cite{braun2014ofdm}

\vspace{-5mm}
\begin{align}
    \zeta(t)=\sum_{k=1}^K\alpha_k x(t-\tau_k)e^{j2\pi f_{d,k}t} + q(t)
    \label{y_time}
\end{align}
where $\alpha_k$ denotes the attenuation factor of the target $k$, $f_{d,k}$ denotes the Doppler frequency of the $k$-th target, $\tau_k$ is the target $k$ delay, $q(t) \in \mathbb{C}$ is the complex additive white Gaussian noise (CAWGN) with zero mean and variance of $2\sigma^2$. 

Conventional OFDM receivers typically involve signal sampling at each symbol interval, followed by fast Fourier transform (FFT) processing to extract the modulated symbols. In typical situations, delay and Doppler effects are compensated and inter-symbol interference (ISI) is addressed. The cyclic prefix is commonly used to reduce inter-symbol interference that can occur due to multipath propagation. However, in our specific scenario, the path delay and Doppler effects may deviate from the expected range due to significant uncertainties regarding the target's position and speed. Hence, ISI may occur in both the time and frequency domains. Accordingly, the extracted received samples after FFT can be expressed as:

\vspace*{-5mm}
\begin{align}
    y(m,n) =&  \sum_{k=1}^K\alpha_k e^{j2\pi n T_0 f_{d,k}} e^{-j2\pi m \Delta f \tau_k} (\sqrt{\gamma_c} s_c(m,n) \notag\\
    &+ \sqrt{\gamma_s} s_s(m,n)) + \text{ISI} + u(m,n)
    \label{y_freq}
\end{align}
where $\text{ISI}$ represents the inter-symbol interference resulting from Doppler and delay effects. $u(m,n) \in \mathbb{C}$ denotes the CAWGN with a mean of zero and a variance of $2\sigma^2$ on the $m$-th subcarrier and the $n$-th OFDM symbol, obtained from sampling and FFT over $\zeta(t)$.

\subsection{Range and Velocity Estimation}\label{est}
Several methods are described in \cite{braun2014ofdm} for estimating range and velocity. In this work, we focus on a computationally efficient method, which unfolds in two stages: an initial estimation of range, followed by Doppler estimation after delay compensation. The time-of-flight (ToF) from the transmitter to the target and then to the receiver can be derived from range assessment utilizing the periodogram \cite{pucci2021performance}. To achieve this, we eliminate the transmitted sensing symbols from the received echoes via point-wise division, expressed as:

\vspace*{-3mm}
\begin{align}
    g(m,n) =    
    \begin{cases}
      \frac{y(m,n)}{s_s(m,n)} & \text{   , If  } s_s(m,n) \neq 0\\
      0 & \text{   , If  } s_s(m,n) = 0
    \end{cases}\label{gmn}
\end{align}

Then, we perform $M$-point inverse fast Fourier transform (IFFT) on the $n$-th column of the $g(m,n)$ of PRS as 

\vspace*{-5mm}
\begin{align}
    r_n(l) =& |\text{IFFT}(g(m,n))| = |\sum_{k=1}^K (\alpha_k e^{j2\pi n T_0 f_{d,k}} \sum_{m=0}^{M-1} (\sqrt{\gamma_s} \notag \\ 
    &+ \sqrt{\gamma_c}\frac{s_c(m,n)}{s_s(m,n)} ) e^{-j2\pi m \Delta f  \frac{R^{tot}_k}{c_0}} e^{j2\pi\frac{ml}{M}}\notag \\
    &+ \frac{u'(m,n)}{s_s(m,n)}e^{j2\pi\frac{ml}{M}})|
    \label{r_l}
\end{align}
where $l= 0,...,M-1$ and $|.|$ is the absolute value and $u'(m,n) = u(m,n) + \text{ISI}$. The absolute value is taken to reduce Doppler sensitivity. We replaced $\tau_k$ by $\frac{R^{tot}_k}{c_0}$, where $R^{tot}_k$ denotes the bistatic range, \emph{i.e.} the distance from the gNB to the target and then to the receiver, and $c_0$ represents the speed of light. Since data symbols are random, $\frac{s_c(m,n)}{s_s(m,n)}$ would be analogous to noise, 
so the maximum value occurs when the arguments of $e^{-j2\pi m \Delta f  \frac{R^{tot}_k}{c_0}} e^{j2\pi\frac{ml}{M}}$ in \eqref{r_l} cancel each other. Then, we obtain IFFT over all columns of the $g(m,n)$ and average over them to increase the SNR as follows

\vspace*{-5mm}
\begin{align}
   \overline{\rm r}(l) = \frac{1}{N}\sum_{n=0}^{N-1} r_n(l)
    \label{r_l2}
\end{align}

Afterward, the index of the maximum value, denoted as $\hat{l}_k$ in \eqref{r_l2}, needs to be determined. Then, the bistatic distance of each target can be computed as \cite{wei20225g} 

\vspace*{-2mm}
\begin{equation}
    \hat{R}^{tot}_k = \frac{\hat{l}_kc_0}{\Delta f M}
    \label{r_tot}
\end{equation}

The resolution of the range can be determined as follows

\vspace*{-3mm}
\begin{equation}
    \Delta R = \frac{c_0}{\Delta f M}
    \label{R_res}
\end{equation}

The maximum detection range can be expressed as follows

\vspace*{-2mm}
\begin{equation}
    R_{max} = \frac{c_0 M}{\Delta f M} = \frac{c_0}{\Delta f}
    \label{R_MAX}
\end{equation}

Once the delay is compensated, to estimate Doppler, we compute the FFT with $N$ points on the $m$-th row of the matrix $g(m,n)$ as follows

\vspace*{-5mm}
\begin{align}
    v_m(d) =& |\text{FFT}(g(m,n))| = |\sum_{k=1}^K (\alpha_k \sum_{n=0}^{N-1}(\sqrt{\gamma_s} \notag \\ 
    &+ \sqrt{\gamma_c}\frac{s_c(m,n)}{s_s(m,n)} ). e^{j2\pi n T_0 f_{d,k}} e^{-j2\pi\frac{nd}{N}} \notag \\ 
    &+ \frac{u'(m,n)}{s_s(m,n)}e^{-j2\pi\frac{nd}{N}})|
    \label{v_m}
\end{align}
where $d= 0,...,N-1$. Next, we perform FFT over all rows of the matrix $g(m,n)$ and then average the results as follows

\vspace*{-4mm}
\begin{align}
   \overline{\rm v}(d) = \frac{1}{M}\sum_{m=0}^{M-1} v_m(d)
    \label{v_m2}
\end{align}

Afterward, we determine the index of the maximum value, denoted as $\hat{d}_k$, for target $k$ in equation \eqref{v_m2}, and estimate the Doppler frequency as \cite{pucci2021performance} 

\vspace*{-3mm}
\begin{equation}
    \hat{f}_{d,k} = \frac{\hat{d}_k}{T_s N}
\end{equation}
 
The velocity can be obtained from Doppler frequency as follows

\vspace*{-4mm}
\begin{equation}
    v_k = \frac{c_0 f_{d,k}}{2 f_c}
\end{equation} 
 where $f_c$ is the central frequency. Consequently, velocity can be estimated as

\vspace*{-4mm}
\begin{equation}
    \hat{v}_k = \frac{\hat{d}_kc_0}{2T_s f_c N}
\end{equation}

The resolution of velocity  can be expressed as 

\vspace*{-3mm}
\begin{equation}
    \Delta\hat{v} = \frac{c_0}{2T_s f_c N}
\end{equation}
and the maximum detectable velocity estimation can be expressed as follows 

\vspace*{-3mm}
\begin{equation}
    \hat{v}_{max} = \frac{c_0}{2T_s f_c}
    \label{v_max}
\end{equation}

\begin{figure}[!t]
\centering
\mbox{\includegraphics[width=\linewidth]{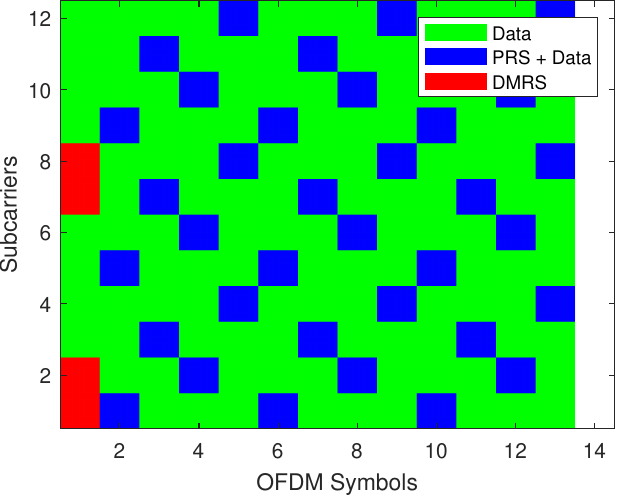}}
\caption{Resource allocation between PRS and PDSCH 
 showing a slot and a PRB of the whole resource grid with the comb size equal to 4.}
\label{resourcegrid}
\end{figure}

\begin{table}[!t]
\caption{Range interval between ambiguities for regular PRS pattern (meter).}\label{table3}
\centering
\begin{tabular}{ |p{2cm}|p{1cm}|p{1cm}|p{1cm}|p{1cm}|  }
 \hline
 & comb 2 & comb 4 & comb 6 & comb 12\\
 \hline 
 SCS = 15 kHz & 4996 & 2498 & 1665 & 832\\ \hline
 SCS = 30 kHz & 2498 & 1249 & 832 & 416 \\ \hline
 SCS = 60 kHz & 1250 & 624 & 416 & 208 \\ \hline
 SCS = 120 kHz & 624 & 312 & 208 & 104\\ \hline
 SCS = 240 kHz & 312 & 156 & 104 & 52\\ \hline
\end{tabular}
\end{table}

\subsection{Range Ambiguity and Ghost Targets}\label{sec:ambiguity}
Originally, PRS was designed for positioning in which time-frequency uncertainties are addressed by the time-frequency servo loop. This is not the case for sensing, as the uncertainty ranges are potentially larger.
The periodic structure of PRS can be adjusted by varying the comb size, affecting the layout of the resource grid. This periodicity, when utilizing the range estimation algorithm discussed in Section \ref{est}, leads to the emergence of ghost targets. Although it enhances the maximum detectable range and velocity compared to \cite{wei20225g}, the presence of ghost targets poses a challenge in distinguishing between real and false targets, as can be seen in Figs. \ref{fig:comb4a}, \ref{fig:comb4c}, \ref{fig:comb12a} and \ref{fig:comb12c}. Table \ref{table3} provides the periodic distances among ghost targets, calculated from the two-way propagation delay using Eq. \eqref{R_amb}, with SCS representing the subcarrier spacing. Notably, for one-way propagation delay, ambiguity intervals are double the values presented in Table \ref{table3}.

\vspace*{-2mm}
\begin{equation}
    \Delta R_{ambiguity} = \frac{c_0}{2 K_{comb}^{PRS}\Delta f}
    \label{R_amb}
\end{equation}

In \eqref{R_amb}, $K_{comb}^{PRS}$ represents the PRS comb size. Hence, to correctly estimate the range of targets without encountering ambiguity, it is essential to adopt a pragmatic approach in accordance with 3GPP standards.

\section{Spectral efficiency enhancement and ghost target elimination}\label{Sec5}
In this section, we explain the superposition of PDSCH for data transmission and PRS for sensing within an OFDM resource grid while the sensing coverage range is increased in comparison with \cite{wei20225g} without encountering ghost targets. We also explain the effect of power allocation in superposition between PDSCH and PRS. 

\subsection{Superposition of PDSCH and PRS and Interference Cancellation}
The desirable anti-noise and auto-correlation characteristics of PRS indicated the fact that PRS would still have acceptable performance in low SNR or, equivalently, high interference. This would lead to the idea of the superposition of PRS and PDSCH while still expecting to have good performance for range and velocity estimation using the periodogram explained in \ref{est} with proper power allocation. On the other hand, to remove the interference caused by PRS for data decoding, we need to estimate the channel using DMRS. To be able to estimate the channel, we need to configure the PRS, PDSCH and DMRS so that PRS and DMRS symbols do not have any superpositions. The better accuracy in channel estimation results in more effective interference cancellation. This configuration is explained in section \ref{results}. Once we obtain the estimated channel, i.e., $\hat{H}(m,n)$, we try to reconstruct the received PRS signal without PDSCH as follows

\vspace*{-4mm}
\begin{align}
    \hat{y}(m,n) = \hat{H}(m,n) \sqrt{\gamma_s} s_s(m,n)
\end{align}

Then, we remove the estimated received PRS from the actual received signal to reduce the interference for PDSCH caused by PRS as follows

\vspace*{-4mm}
\begin{align}
    \Tilde{y}(m,n) = y(m,n) - \hat{y}(m,n)
\end{align}

Eventually, we can use $\Tilde{y}(m,n)$ for data decoding. In Sec. \ref{alg1} and \ref{alg2}, we introduce two algorithms to remove the ghost targets without reducing range and velocity resolution. Unlike the approach in \cite{wei20225g}, our algorithms maintain the maximum detectable range.

\subsection{Algorithm 1 : Comb Size is Equal to $2$ or $4$ \label{alg1}}
Initially, DMRS is extracted from the received slots. Then, we independently apply the range estimation algorithm described in Section \ref{est} on DMRS and PRS. We define $\overline{\rm r}_{PRS}$ and $\overline{\rm r}_{DMRS}$ as the result of IFFT from PRS and DMRS, respectively similar to \eqref{r_l2} with the same IFFT size. Afterward, we find the index of the maximum value of

\vspace*{-2mm}
\begin{equation}
    \overline{\rm r}_{PRS,DMRS} = \overline{\rm r}_{PRS} \odot  \overline{\rm r}_{DMRS}
    \label{eq:alg1}
\end{equation}
where 
$\odot$ stands for Hadamard product. This technique preserves the range resolution and maximum unambiguous range as derived in \eqref{R_res} and \eqref{R_MAX}, respectively. Ghost targets resulting from range estimation using DMRS appear in positions where PRS with comb sizes of $2$ and $4$ do not exhibit any ghost targets, and vice versa. By performing element-wise multiplication of $\overline{\rm r}_{PRS}$ and $\overline{\rm r}_{DMRS}$, we effectively eliminate the ghost targets, facilitating clear differentiation between real targets and ghost ones.

\subsection{Algorithm 2 : Comb Size is Equal to $6$ or $12$\label{alg2}}
When the comb size is either $6$ or $12$, some ghost targets from PRS and DMRS coincide by employing Algorithm $1$. Therefore, we need another method for comb sizes $6$ and $12$ to remove the ghost targets. In this case, from the extracted PRS and DMRS of the received PDSCH slots, we can derive $g_{DMRS}(m,n)$ and $g_{PRS}(m,n)$ as explained in \eqref{gmn}. Next, to increase SNR, the sum over all columns of $s$-th slot is computed as follows

\vspace*{-1mm}
\begin{align}
    g_{PRS}^{tot,s}(m) = \sum_{n =1}^{N'-1} g_{PRS}^{s}(m,n)
    \label{eq:g_{PRS}^{tot,s}}
\end{align}
where $N'$ is the number of PRS symbols in a time slot and $g_{PRS}^{s}(m,n)$ is extracted from $s$-th slot of $g_{PRS}(m,n)$. $n$ in \eqref{eq:g_{PRS}^{tot,s}} starts from $1$ because of PRS configuration explained in Sec. \ref{results}. Then, $g_{PRS}^{tot,s}(m)$ and $g_{DMRS}^{s}(m,1)$ are normalized over the rows to regularize their impact in the IFFT as

\vspace*{-1mm}
\begin{align}
    \Tilde{g}_{PRS}^{tot,s}(m) &= \frac{g_{PRS}^{tot,s}(m)}{\text{max}_{m}\{g_{PRS}^{tot,s}(m)\}}\\
    \Tilde{g}_{DMRS}^{s}(m) &= \frac{g_{DMRS}^{s}(m,1)}{\text{max}_{m}\{g_{DMRS}^{s}(m,1)\}}
\end{align}
where $g_{DMRS}^{s}(m,1)$ is extracted from $s$-th slot of $g_{DMRS}(m,n)$. The reason that $n=1$ for $g_{DMRS}^{s}$ is because of DMRS and PDSCH configuration specified in Sec. \ref{results}.
Following that, an $M$-point IFFT is performed on the summation of $\Tilde{g}_{PRS}^{tot,s}(m)$ and $\Tilde{g}_{DMRS}^{s}(m)$ as shown below

\begin{equation}
    \kappa_{s}(l) = |\text{IFFT}(\Tilde{g}_{PRS}^{tot,s}(m) + \Tilde{g}^s_{DMRS}(m))|
    \label{eq:ifftalg2}
\end{equation}

Then, we obtain the summation over all OFDM slots, i.e., $S$, to further increase the SNR as follows

\vspace*{-4mm}
\begin{align}
    \overline{\rm \kappa}(l) = \sum_{s =1}^{S} \kappa_{s}(l)
\end{align}

Ultimately, the index of the maximum value of $\overline{\rm \kappa}(l)$ is determined to estimate the bistatic range of targets. The remaining parameters, including bistatic distance, range resolution, and the maximum detection range, can be computed utilizing equations \eqref{r_tot}, \eqref{R_res}, and \eqref{R_MAX}, respectively.
By doing so, we disrupt the periodicity of the PRS and DMRS patterns, thereby suppressing the magnitude of the ghost target peaks. Consequently, distinguishing between real and fake targets becomes straightforward. 

\subsection{Effect of Power Allocation}

Power allocation plays a crucial role not only in ambiguity elimination but also in BER and throughput performance. Increasing the power of PRS for sensing amplifies the effect of $\overline{\rm r}_{PRS}$ in \eqref{eq:alg1} and $\Tilde{g}_{PRS}^{tot,s}(m)$ in \eqref{eq:ifftalg2}, which reduces throughput. Similarly, increasing the power of PDSCH for communication enhances the impact of $\overline{\rm r}_{DMRS}$ in \eqref{eq:alg1} and $\Tilde{g}^s_{DMRS}(m))$ in \eqref{eq:ifftalg2}, thereby increasing throughput. However, allocating excessive power to sensing diminishes the influence of DMRS, while over-allocating power to communication minimizes the effect of PRS in the IFFT process. Therefore, power allocation between sensing and communication should be done based on the number of targets, PRS comb size, and communication requirements. 
Increasing the PRS comb size or the number of targets raises the number of ghost targets. Additionally, achieving low BER requires more communication power, which increases the magnitude of fake peaks after IFFT, making their suppression more challenging. In the following section, we evaluate these effects through simulation results.

\section{Simulation Results \label{results}}
To validate our approach, we exploit MATLAB 5G toolboxes, ensuring compatibility with 3GPP standards. We conduct simulations for a scenario where the gNB and receiver are equipped with $8\times 8$ antennas, two targets are located at a bistatic distance of $711$m and $846$m and moving at a velocities of $2$m/s and $10$m/s. The central frequency is set to $25$GHz with a subcarrier spacing of $120$kHz, and SNR is $15$dB. PRS number of symbols is $12$, PRS starting symbol is $1$, DMRS configuration type is $2$ with $1$ DMRS symbol position in a slot, PDSCH mapping type is "B", the code rate used to calculate transport block sizes is $490/1024$ based on \cite{3gpp2018nr1}. Virtual resources block (VRB) bundle size is set to $4$, and VRB to PRB interleaving is disabled. $\alpha_1 =\alpha_2=1$, PDSCH symbol allocation is $[0,...,12]$, number of layer is $1$ and modulation is QPSK. The low-density parity check (LDPC) decoding algorithm used in the receiver is the normalized min-sum decoding algorithm. 

\begin{figure}[!t]
    \centering
    \begin{subfigure}{0.45\columnwidth} 
        \centering
        \includegraphics[width=\linewidth]{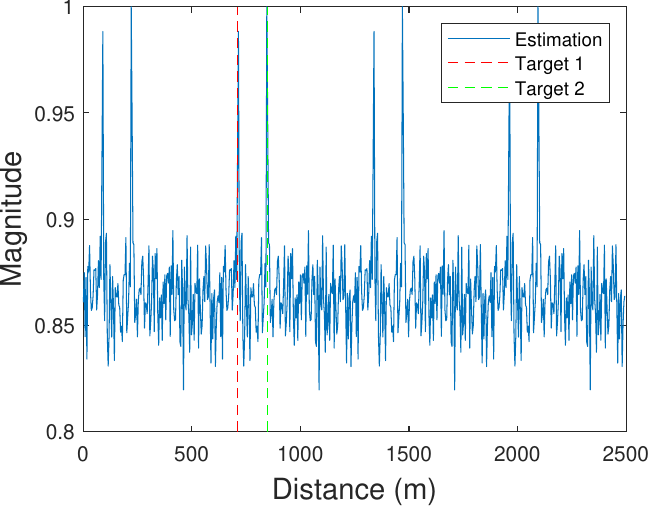}
        \caption{PRS only and $\sqrt{\gamma_s}  = 0.1$}
        \label{fig:comb4a}
    \end{subfigure}
    \vspace{0.1cm}
    \begin{subfigure}{0.45\columnwidth}
        \centering
        \includegraphics[width=\linewidth]{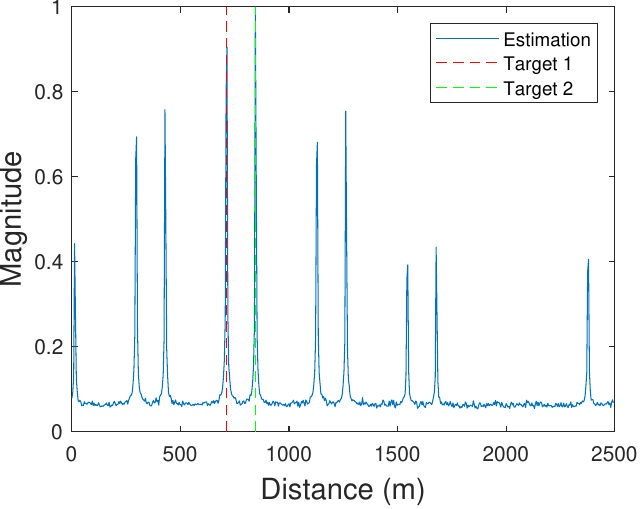}
        \caption{PRS-DMRS, $\sqrt{\gamma_s}  = 0.1$}
        \label{fig:comb4b}
    \end{subfigure}\vspace{0.1cm}
    \begin{subfigure}{0.45\columnwidth}
        \centering
        \includegraphics[width=\linewidth]{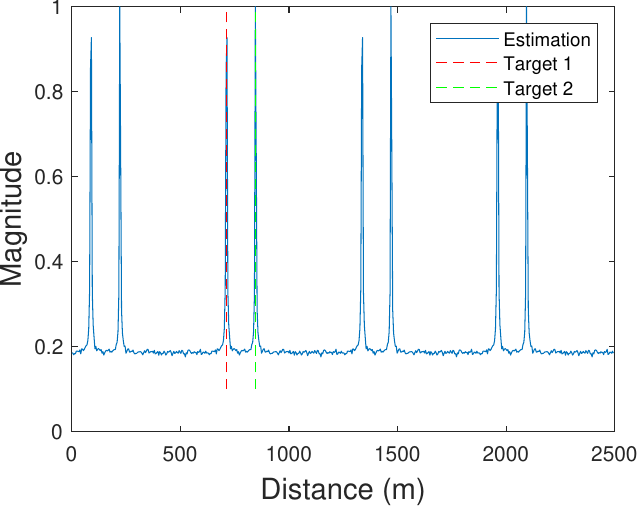}
        \caption{PRS only and $\sqrt{\gamma_s} = 0.5$}
        \label{fig:comb4c}
    \end{subfigure}
    \begin{subfigure}{0.45\columnwidth}
        \centering
        \includegraphics[width=\linewidth]{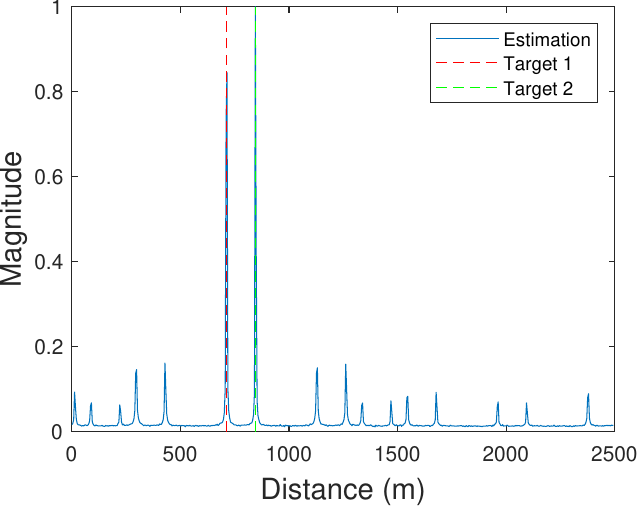}
        \caption{PRS-DMRS, $\sqrt{\gamma_s} = 0.5$}
        \label{fig:comb4d}
    \end{subfigure}
    \caption{Range ambiguity elimination with the presence of $2$ targets with $K_{comb}^{PRS} = 4$,  $f_c=25$GHz and SCS=$120$kHz.}
    \label{fig:comb4}
\end{figure}

\begin{figure}[!t]
    \centering
    \begin{subfigure}{0.45\columnwidth}
        \centering
        \includegraphics[width=\linewidth]{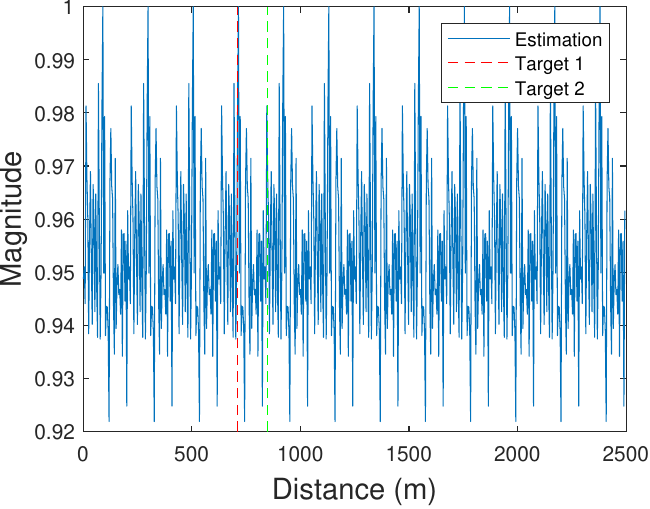}
        \caption{PRS only and $\sqrt{\gamma_s} = 0.1$}
        \label{fig:comb12a}
    \end{subfigure}\vspace{0.1cm}
    \begin{subfigure}{0.45\columnwidth}
        \centering
        \includegraphics[width=\linewidth]{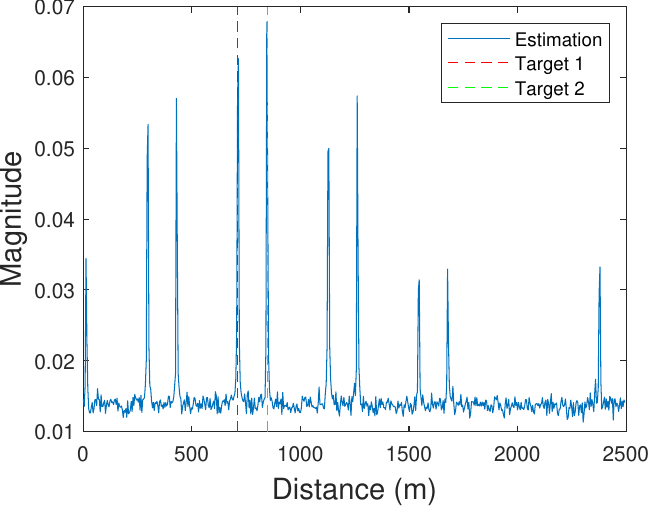}
        \caption{PRS-DMRS, $\sqrt{\gamma_s} = 0.1$}
        \label{fig:comb12b}
    \end{subfigure}\vspace{0.1cm}
    \begin{subfigure}{0.45\columnwidth}
        \centering
        \includegraphics[width=\linewidth]{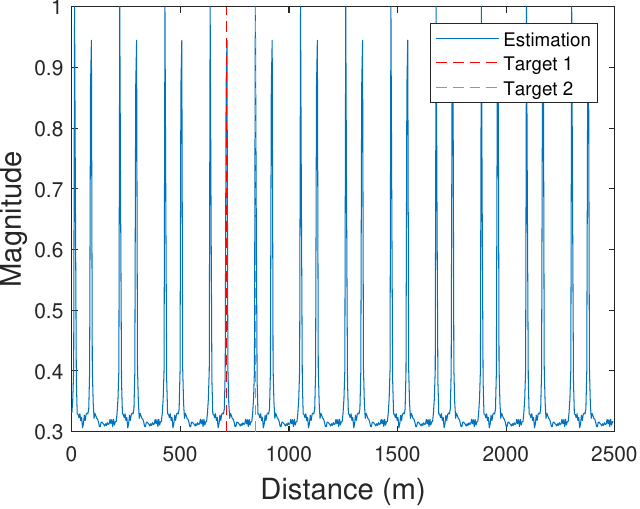}
        \caption{PRS only and $\sqrt{\gamma_s} = 0.5$}
        \label{fig:comb12c}
    \end{subfigure}
    \begin{subfigure}{0.45\columnwidth}
        \centering
        \includegraphics[width=\linewidth]{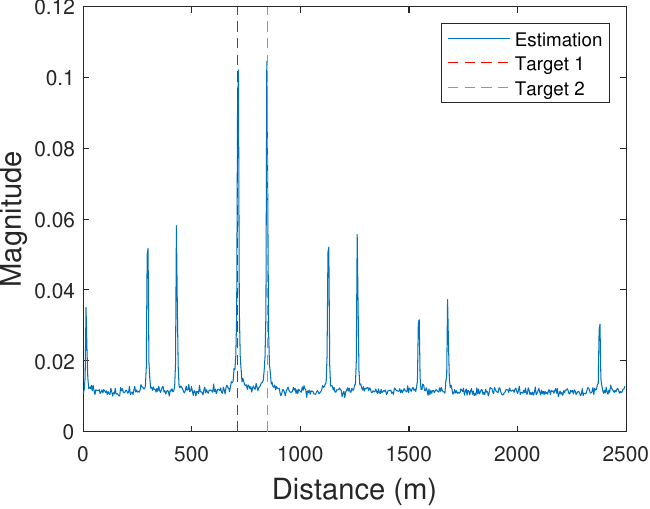}
        \caption{PRS-DMRS, $\sqrt{\gamma_s} = 0.5$}
        \label{fig:comb12d}
    \end{subfigure}
    \caption{Range ambiguity elimination with the presence of 2 targets with $K_{comb}^{PRS} = 12$,  $f_c=25$GHz and SCS=$120$kHz.}
    \label{fig:comb12}
\end{figure}

Figures \ref{fig:comb4a} and \ref{fig:comb4c} illustrate the ambiguity in range estimation when the PRS comb size is $4$. As shown, this ambiguity makes distinguishing between real and ghost targets quite challenging. When $\sqrt{\gamma_s} = 0.1$ and the power allocated to PRS is low, the combination of PRS and DMRS fails to effectively mitigate the ambiguity, as depicted in Fig. \ref{fig:comb4b} using Algorithm 1. However, with appropriate power allocation between PRS and PDSCH, ambiguity can be solved using Algorithm 1. For instance, when $\sqrt{\gamma_s}  = 0.5$ and PRS comb size is $4$, as shown in Fig. \ref{fig:comb4d}, the real targets are distinguishable. It should be noted that the magnitudes in Figures \ref{fig:comb4} are normalized. As the comb size increases, the ambiguity becomes more severe, as depicted in Figures \ref{fig:comb12a} and \ref{fig:comb12c}. With $\sqrt{\gamma_s} = 0.1$, the effect of DMRS dominates over PRS due to the higher power allocation to PDSCH. Therefore, the range estimation with PRS and DMRS combined using algorithm 2 does not remove the range ambiguity, as shown in Fig. \ref{fig:comb12b}. Conversely, utilizing Algorithm 2 and allocating sufficient power to PRS significantly suppresses the magnitude of fake peaks in range estimation, enabling clear differentiation between real and ghost targets when PRS comb size is $12$, as illustrated in Fig. \ref{fig:comb12d}. As a result, we can accurately identify the two highest peaks and estimate the ranges of the targets. It is worth mentioning that the number of targets can be determined using model order selection based on information-theoretic criteria, as outlined in \cite{mariani2015model}.

\begin{figure}[!t]
\centering
\mbox{\includegraphics[width=\linewidth]{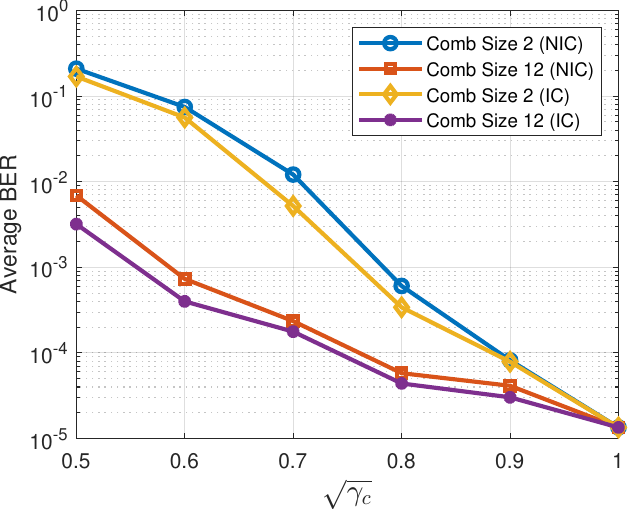}}
\caption{Impact of power allocation on BER for PRS comb sizes 2 and 12.}
\label{fig:BER}
\end{figure}

\begin{figure}[!t]
\centering
\mbox{\includegraphics[width=\linewidth]{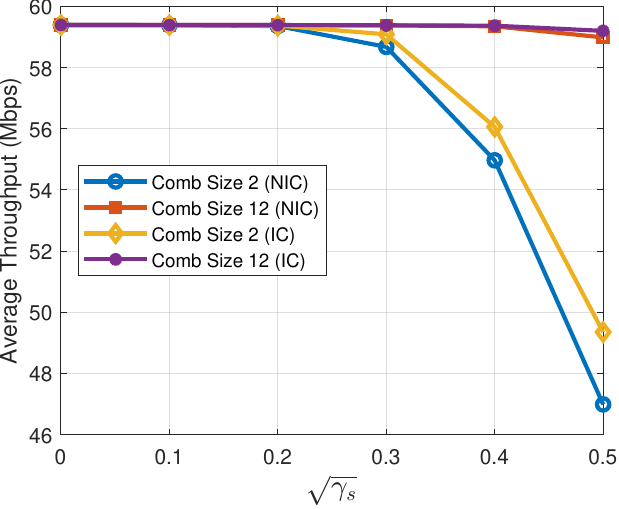}}
\caption{Impact of power allocation on throughput for PRS comb sizes 2 and 12.}
\label{fig:BR}
\end{figure}

In Figures \ref{fig:BER} and \ref{fig:BR}, the influence of power allocation on average BER and throughput is evident when varying $\gamma_c$ and $\gamma_s$ with different comb sizes. NIC and IC in Figures \ref{fig:BER} and \ref{fig:BR} indicate no interference cancellation and with interference cancellation, respectively. It can be seen that as comb size decreases, the interference between PRS and PDSCH increases, resulting in higher BER and lower throughput. Moreover, we notice from Figures \ref{fig:BER} and \ref{fig:BR}, the proposed interference cancellation method significantly improves the BER and throughput. As an example, when PRS comb size is $2$ and $\sqrt{\gamma_c} = 0.7$, the BER with interference cancellation is $0.0052$ compared to $0.0121$ without interference cancellation, demonstrating an improvement of nearly $57\%$. Indeed, PRS with comb sizes of 2 and 12 create the maximum and minimum interference for PDSCH, resulting in the lower and upper bounds for throughput, or the upper and lower bounds for BER, respectively. In this simulation, we employed practical channel estimation using DMRS rather than perfect channel estimation. Notably, using perfect channel estimation in the proposed interference cancellation method would further enhance communication performance in terms of BER and throughput. It should be noted that the distance of the target does not affect throughput and BER due to the synchronization performed at the receiver using DMRS.

\section{Conclusion}\label{Sec6}
In this work, we proposed the superposition of existing 5G NR signals in communication and positioning, namely PDSCH and PRS, in the ISAC framework to enhance spectral efficiency. By allocating full resources in both the time and frequency domains to PRS and PDSCH, combined with appropriate power allocation, we aimed to maximize sensing resolution for range and velocity estimation while maintaining communication performance. To achieve this, we introduced a novel algorithm to mitigate the interference caused by PRS during data decoding, leading to a reduction in BER by up to $57\%$ and throughput enhancement. Moreover, we proposed two novel algorithms for different PRS comb sizes to mitigate range estimation ambiguity and remove the ghost targets by joint exploitation of PRS and DMRS available in PDSCH while the maximum detection range is enhanced. Eventually, we validated our methods using the MATLAB 5G toolbox, ensuring compliance with 3GPP standards. Our study provides the first proof of concept for the superposition of PRS and PDSCH symbols in an OFDM resource grid, taking 3GPP standard constraints into account. 
For future work, further investigation into power allocation under varying SNRs and its impact on BER and throughput, as well as the trade-off between detection accuracy and BER/throughput with different subcarrier spacing, will be of interest.

\bibliographystyle{IEEEtran}
\bibliography{ref}

\end{document}